\documentstyle[aps,epsf,prbbib]{revtex}

\newcommand{\mybold}[1]{\text{\normalsize\boldmath{$#1$}}}
\newcommand{\mycite}[2]{#1\cite{#2}}
\begin{document}
\wideabs{
  \title{Correlated hopping of electrons:
    Effect on the Brinkman-Rice transition and
    the stability of metallic ferromagnetism}
  \author{M.\ Kollar$^*$ and D.\ Vollhardt}
  \date{July 31, 2000}
  \address{Theoretical Physics III,
    Center for Electronic Correlations and Magnetism, Institute of
    Physics, University of Augsburg, D-86135 Augsburg, Germany}
  \maketitle
  \begin{abstract}
    We study the Hubbard model with bond-charge interaction
    (``correlated hopping'') in terms of the Gutzwiller wave function.
    We show how to express the Gutzwiller expectation value of the
    bond-charge interaction in terms of the correlated momentum-space
    occupation. This relation is valid in all spatial dimensions. We
    find that in infinite dimensions, where the Gutzwiller
    approximation becomes exact, the bond-charge interaction lowers
    the critical Hubbard interaction for the Brinkman-Rice
    metal-insulator transition. The bond-charge interaction also
    favors ferromagnetic transitions, especially if the density of
    states is not symmetric and has a large spectral weight below the
    Fermi energy.\nocite{address}
  \end{abstract}
}

  The microscopic origin of magnetic ordering in systems like
  transition metals, transition-metal oxides, and high-temperature
  superconductors is intricate, since it is due to correlations between
  the electrons. The simplest model to attempt a description of such
  systems is the single-band Hubbard
  model\mycite{}{Hubbard63a,Kanamori63a,Gutzwiller63a+64a+65a}
  \begin{eqnarray}
    \hat{H}_{\text{Hubbard}}
    &=&
    \sum_{ij\sigma}
    t_{ij}
    \hat{c}_{i\sigma}^{+}
    \hat{c}_{j\sigma}^{\phantom{+}}
    +
    U
    \sum_{i}\hat{n}_{i\uparrow}\hat{n}_{i\downarrow}
    ,\label{u-hamiltonian}
  \end{eqnarray}
  where hats indicate operators.  This model describes the competition
  between kinetic and potential energy which is at the heart of the
  quantum-mechanical correlation problem\mycite{.}{Gebhard97a} The
  Hubbard interaction $U$ represents the Coulomb repulsion of
  electrons in the same orbital at a given lattice site. It is given
  by the matrix element $U$ $=$ $\langle
  ii|V(\mybold{r}-\mybold{r'})|ii\rangle$ of the Coulomb potential,
  and is typically on the order of a few eV.  The matrix elements
  involving neighboring lattice sites $i$ and $j$ are generally
  smaller than $U$, but may not be negligibly small. One of them is
  the bond-charge interaction $X_{ij}$ $=$ $\langle ii|$
  $V(\mybold{r}-\mybold{r'})$ $|ij\rangle$\mycite{,}{Hubbard63a,%
    Gammel88a+Campbell88a,Hirsch89a+89b+90a+91a,footnote1} which is
  typically on the order of 0.1-1 eV and hence is comparable in
  magnitude to the tight-binding hopping amplitude $t_{ij}$. It
  describes a density-dependent hopping of the electrons
  \begin{eqnarray}
    \hat{H}_X
    &=&
    \sum_{ij\sigma}
    X_{ij}
    \hat{c}_{i\sigma}^{+}
    \hat{c}_{j\sigma}^{\phantom{+}}
    (\hat{n}_{i\bar{\sigma}}+\hat{n}_{j\bar{\sigma}})
    ,\label{x-hamiltonian}
  \end{eqnarray}
  which only contributes if the lattice site from or onto which an
  electron with spin $\sigma$ is hopping is occupied by an electron
  with spin $\bar{\sigma}$ (``correlated hopping'').  The effect of
  $\hat{H}_X$ competes strongly with both the kinetic energy and the
  Hubbard interaction.  Correlated hopping of spin-$\sigma$ electrons
  between two sites is enhanced if spin-$\bar{\sigma}$ electrons are
  present, but this in turn will cost the latter kinetic energy, as
  well as Coulomb energy for the double occupations.  Moreover, in a
  band picture the coupling of densities and kinetic energy can lead
  to a band narrowing which lowers the amount of energy that is
  necessary for a ferromagnetic spin polarization.  These mechanisms
  explain why, in principle, $\hat{H}_X$ may play an important role in
  the stabilization of ferromagnetism and the localization of
  electrons.

  The model $\hat{H}$ $=$ $\hat{H}_{\text{Hubbard}}$ $+$ $\hat{H}_{X}$
  cannot be solved exactly, and only mean-field and finite-size
  diagonalization results are available\mycite{.}{Amadon96a} Recently,
  Schiller\mycite{}{Schiller99a} showed how to incorporate
  $\hat{H}_{X}$ into the framework of dynamical mean-field
  theory\mycite{,}{Metzner89a,Georges96a} but no numerical results
  have yet been obtained for $\hat{H}$.
  
  One of the standard tools to approach the correlated electron
  problem is the Gutzwiller wave
  function\mycite{}{Gutzwiller63a+64a+65a}
  \begin{equation}
    |\Psi_{\text{G}}\rangle
    =
    \prod_i\,
    [1-(1-g)\hat{n}_{i\uparrow}\hat{n}_{i\downarrow}]\,
    |\Phi_{0}\rangle
    ,
  \end{equation}
  where $g$ is a variational parameter ($0\leq g\leq 1$), and the
  starting wave function $|\Phi_{0}\rangle $ is a product state of
  spin-up and spin-down Fermi seas. By construction both
  $|\Phi_{0}\rangle $ and $|\Psi_{\text{G}}\rangle $ are
  translationally invariant and have a fixed particle density $n$ $=$
  $n_{\uparrow}$ $+$ $n_{\downarrow}$ and magnetization $m$ $=$
  $n_{\uparrow}$ $-$ $n_{\downarrow}$.  The uncorrelated case
  $U=X_{ij}=0$ corresponds to $g=1$, while $U=\infty $ forbids any
  doubly occupied sites and thus corresponds to $g=0$. For $m$ $\neq$
  $0$ and $g\neq 0$ the wave function $|\Psi_{\text{G}}\rangle$
  describes an itinerant ferromagnetic state.  Starting wave functions
  $|\Phi_{0}\rangle$ with other broken symmetries can also be
  considered. Here, however, we will only consider paramagnetism and
  ferromagnetism.
  
  Using the Gutzwiller wave function one may, in principle, calculate
  expectation values of any operator $\hat{A}$ as $\langle
  \hat{A}\rangle_{\text{G}}$ $=$ $\langle
  \Psi_{\text{G}}|\hat{A}|\Psi_{\text{G}}\rangle /\langle
  \Psi_{\text{G}}|\Psi_{\text{G}}\rangle $. The energy expectation
  value $E$ $=$ $\langle \hat{H}\rangle_{\text{G}}$, when optimized
  with respect to $g$, is an upper bound for the exact ground-state
  energy of $\hat{H}$ by the variational principle. The variational
  energy $E$ can be written as
  \begin{equation}
    E=
    \langle\hat{H}\rangle_{\text{G}}
    =
    \sum_{\mybold{k}\sigma}
    \epsilon_{\mybold{k}}
    n_{\mybold{k}\sigma}^{\phantom{0}}
    +
    \sum_{\mybold{k}\sigma}
    2\xi_{\mybold{k}}
    \text{Re}(x_{\mybold{k}\sigma}^{\phantom{0}})
    +
    U\,d
    ,
  \end{equation}
  where the $k$-space occupation $n_{\mybold{k}\sigma}^{\phantom{0}}$,
  bond-charge occupation $x_{\mybold{k}\sigma}^{\phantom{0}}$, and
  double occupation $d$ are defined by ($L$ is number of lattice sites)
  \begin{eqnarray}
    n_{\mybold{k}\sigma}^{\phantom{0}}
    &=&
    \frac{1}{L}\sum_{i\neq j}
    e^{i\mybold{k}(\mybold{R}_i-\mybold{R}_j)}
    \langle
    \hat{c}_{i\sigma}^{+}
    \hat{c}_{j\sigma}^{\phantom{+}}
    \rangle_{\text{G}}
    ,\label{kocc}
    \\
    x_{\mybold{k}\sigma}^{\phantom{0}}
    &=&
    \frac{1}{L}\sum_{i\neq j}
    e^{i\mybold{k}(\mybold{R}_i-\mybold{R}_j)}
    \langle
    \hat{n}_{i\bar{\sigma}}
    \hat{c}_{i\sigma}^{+}
    \hat{c}_{j\sigma}^{\phantom{+}}
    \rangle_{\text{G}}
    ,\label{xocc}
    \\
    d
    &=&
    \frac{1}{L}\sum_{i}
    \langle
    \hat{n}_{i\uparrow}
    \hat{n}_{i\downarrow}
    \rangle_{\text{G}}
    .\label{docc}
  \end{eqnarray}
  They each depend on $g$, $n$, and $m$. Here $\epsilon_{\mybold{k}}$
  and $\xi_{\mybold{k}}$ are the Fourier transforms of $t_{ij}$ and
  $X_{ij}$, respectively. By convention, $t_{ii}$ $=$ $X_{ii}$ $=$
  $0$.
  
  In general the expectation values in Eqs.~(\ref{kocc})-(\ref{docc})
  are not independent of one another.  In particular, for the
  Gutzwiller wave function the $k$-space and bond-charge occupation
  are closely related. Using the techniques of
  Ref.~\onlinecite{Metzner87a+88a} we obtain, for all Bravais lattices
  in arbitrary dimensions $D$,
  \begin{equation}
    x_{\mybold{k}\sigma}^{\phantom{0}}
    =
    \frac{[(1+g)\,n_{\mybold{k}\sigma}^{0}-g]\,
      n_{\mybold{k}\sigma}^{\phantom{0}}
      -
      n_{\mybold{k}\sigma}^{0}}{1-g}
    +
    n_{\bar{\sigma}}n_{\mybold{k}\sigma}^{0} 
    -
    d
    ,
  \end{equation}
  with $n_{\mybold{k}\sigma}^{0}$ $=$
  $n_{\mybold{k}\sigma}^{\phantom{0}}|_{g=1}$ as the uncorrelated
  Fermi function. It
  should be noted that although $x_{\mybold{k}\sigma}^{\phantom{0}}$
  and $n_{\mybold{k}\sigma}^{\phantom{0}}$ are linearly related, the
  bond-charge energy and the kinetic energy will generally {\em not}
  be, since the regions of the Brillouin zone inside
  ($n_{\mybold{k}\sigma}^{0}=1$) and outside
  ($n_{\mybold{k}\sigma}^{0}=0$) of the Fermi sea contribute
  differently.
  
  {\em Gutzwiller approximation}.--- Now we turn to evaluate the
  variational energy $E$ within the Gutzwiller approximation, which is
  known to yield the exact evaluation of expectation values in terms
  of the Gutzwiller wave function in the limit of infinite spatial
  dimensions ($D$ $\rightarrow$
  $\infty)$\mycite{.}{Metzner89a,Metzner87a+88a,Metzner89b,Gebhard90a}
  It describes a Fermi liquid\mycite{}{Vollhardt84a} with piecewise
  constant $k$-space occupation,
    \begin{equation}
    n_{\mybold{k}\sigma}^{\phantom{0}}
    =
    n_{\sigma} +
    (n_{\mybold{k}\sigma}^{0}-n_{\sigma})q_{\sigma}
    ,
  \end{equation}
  where the
  discontinuity at the Fermi surface is given by
  \begin{equation}
    q_{\sigma}
    =
    \frac{[\sqrt{(n_{\sigma}-d)(1-n+d)}
      +\sqrt{(n_{\bar{\sigma}}-d)\,d}\,]^2}{n_{\sigma}\,(1-n_{\sigma})}
    .
  \end{equation}
  The variational parameter $g$ is related to the double occupation
  $d$ by
  \begin{equation}
    g^{2}
    =
    \frac{(n_{\uparrow}-d)(n_{\downarrow}-d)}{(1-n+d)\,d},
  \end{equation}
  and it is convenient to use the latter
  as variational parameter. For the variational energy we obtain 
  \begin{equation}
    E
    =
    E^{0}-\sum_{\sigma}(1-q_{\sigma})(\epsilon_{0\sigma}+\xi_{0\sigma})
    +\,U_{\text{eff}}\;(d-n_{\uparrow}n_{\downarrow})
    ,\label{energy}
  \end{equation}
  where $\epsilon_{0\sigma}$ $=$
  $(1/L)\sum_{\mybold{k}}\epsilon_{\mybold{k}}n_{\mybold{k}\sigma}^{0}$
  is the uncorrelated kinetic energy, and similarly $\xi_{0\sigma}$
  $=$
  $(1/L)\sum_{\mybold{k}}\xi_{\mybold{k}}n_{\mybold{k}\sigma}^{0}$.
  The effective Hubbard interaction $U_{\text{eff}}$ and the
  uncorrelated variational energy $E^0$ ($=$ $E|_{g=1}$) in
  Eq.~(\ref{energy}) are given by
  \begin{eqnarray}
    U_{\text{eff}}
    &=&
    U
    +
    \sum_{\sigma}
    \frac{(1-2n_{\sigma})}{(1-n_{\sigma})n_{\sigma}}
    \xi_{0\sigma}
    ,
    \\
    E^{0}
    &=&
    \sum_{\sigma}
    (\epsilon_{0\sigma}+2n_{\bar{\sigma}}\xi_{0\sigma})
    +
    U\,n_{\uparrow}n_{\downarrow}
    .
  \end{eqnarray}
  Within the Gutzwiller approximation the bond-charge interaction thus
  leaves the form of the variational energy unchanged, but enters into
  the effective kinetic energy and effective Hubbard interaction via
  $\xi_{0\sigma}$. The effect on the kinetic energy can be interpreted
  as a spin-dependent band narrowing or widening, which is also found
  in the Hartree-Fock approximation of the correlated hopping term and
  can lead to a stabilization of
  ferromagnetism\mycite{.}{Hirsch89a+89b+90a+91a,Amadon96a} However the
  Gutzwiller approximation reveals two distinct effects that cannot be
  resolved in ordinary Hartree-Fock theory, where the suppression of
  double occupancies can only be achieved by spin polarizing the
  system. On the one hand, the kinetic energy increases if
  $\xi_{0\sigma}$ $>$ $0$, i.~e., the effective hopping becomes
  smaller. This corresponds to an effective narrowing of the band,
  which lowers the amount of energy that must be expended for a
  ferromagnetic spin polarization. Furthermore the bond-charge
  interaction contributes to $U_{\text{eff}}$, which suppresses
  double occupancies already at lower values of the bare Hubbard
  interaction $U$.  The balance between these two effects is
  determined by the optimal variational parameter $d$.

  {\em Brinkman-Rice metal-insulator transition}.--- We now discuss
  the effect of $\hat{H}_X$ on the Brinkman-Rice transition that
  occurs in the Gutzwiller approximation at half-filling
  ($n=1$)\mycite{.}{Brinkman70a,Vollhardt84a} For convenience we
  define the strength of the bond-charge interaction $X$ by
  \begin{equation}
    \xi_{0\sigma}
    =
    -X\epsilon_{0\sigma}
    .\label{Xdef}
  \end{equation}
  Note that $\xi_{0\sigma}$ will remain proportional to
  $\epsilon_{0\sigma}$ according to Eq.~(\ref{Xdef}) for all densities
  if $t_{ij}$ and $X_{ij}$ have the same range, e.~g., if they are
  nonzero only for nearest-neighbor sites. In the limit of $D$ $\to$
  $\infty$ both $t_{ij}$ and $X_{ij}$ must both be scaled as
  $1/\sqrt{Z_{ij}}$ ($Z_{ij}$ is the number of neighbors
  $ij$)\mycite{,}{Metzner89a} which is compatible with
  Eq.~(\ref{Xdef}).  The dispersion $\epsilon_{\mybold{k}}$ enters
  only through the density of states (DOS) $N(\epsilon)$ (which
  determines $\epsilon_{0\sigma}$), as expected in dimension
  $D=\infty$. We will consider several densities of states below.
  
  With the above definition of $X$, the Gutzwiller approximation
  energy for $n=1$ and $m=0$ simplifies to $E$ $=$
  $8d(1-2d)(1-X)\epsilon_{0}+U\,d$, where $\epsilon_{0}$ $\equiv$
  $\sum_{\sigma}\epsilon_{0\sigma}$ $<$ $0$.  Optimization with
  respect to $d$ yields a critical value for $U$,
  \begin{equation}
    U_{\text{c}}(X)
    =
    8|\epsilon_{0}|(1-X)
    ,\label{ucx}
  \end{equation}
  above which the localized state with $d=0$ is lowest in energy.
  Hence the Brinkman-Rice transition is moved to lower $U$ for $X>0$,
  i.~e., the bond-charge interaction favors localization.  (Only
  $U\geq 0$ and $X\leq 1$ will be considered from now on.) We find
  that the $U$ dependence of the double occupation $d$, the
  discontinuity of the $k$-space occupation $q$ ($\equiv$
  $q_{\sigma}$), and the energy $E$ is formally the same as in the
  original Brinkman-Rice theory for \mbox{$X$ $=$ $0$}, i.~e.,
  \begin{eqnarray}
    d
    &=&
    \frac{1}{4}\left(1-\frac{U}{U_{\text{c}}}\right)
    ,
    \\
    q
    &=&
    1-\frac{U^{2}}{U_{\text{c}}^{2}}
    ,
    \\
    E
    &=&
    -\frac{U_{\text{c}}}{8}\left(1-\frac{U}{U_{\text{c}}}^{2}\right),
  \end{eqnarray}
  except that $U_{\text{c}}$ now depends on $X$ [Eq.~(\ref{ucx})]. We
  note that the simultaneous vanishing of $q$ and double occupation
  $d$ at a finite value of $U$ is characteristic of the Brinkman-Rice
  transition, in contrast to the numerical solution of the Hubbard
  model, where $d$ remains nonzero across the
  transition\mycite{.}{Georges96a,Schlipf99a}
  
  {\em Ferromagnetic transition.}--- The instability of the
  par\-amagnetic state toward ferromagnetism can be determined from
  the bulk susceptibility $\chi$. For half-filling, we obtain
  \begin{equation}
    \frac{1}{\chi}
    =
    \frac{q}{2N(\epsilon_F)}
    \left[
      (1\!-\!X)\!
      \left(
        1
        \!-\!
        \frac{p\,U(U+2U_{\text{c}})}{(U+U_{\text{c}})^2}
      \right)
      +\frac{r\,X\,U_{\text{c}}}{U+U_{\text{c}}}
    \right]
    ,\label{br-susceptibility}
  \end{equation}
  where we have introduced the dimensionless parameters
  \begin{eqnarray}
    p
    &=&
    4\,N(\epsilon_{F})\,|\epsilon_{0}|
    ,\label{pparam}
    \\
    r
    &=&
    4\,N(\epsilon_{F})\,\epsilon_{F}
    .\label{rparam}
  \end{eqnarray}
  The Fermi energy $\epsilon_{F}$ in Eq.~(\ref{rparam}) represents an
  absolute scale since the first moment of the DOS is fixed at zero
  (due to $t_{ii}=0$). There are two factors in $\chi $ that can
  diverge: either $q\rightarrow 0$, i.~e., the effective band mass
  $m^{\ast}/m=q^{-1}$ diverges at $U_{\text{c}}(X)$, indicating a
  localization transition, or the Stoner-type factor in square
  brackets in Eq.~(\ref{br-susceptibility}) vanishes at
  \begin{equation}
    U_{\text{fm}}(X)
    =
    \frac{\sqrt{r^{2}X^{2}-4p(1-p)(1-X)^{2}}-rX}{2(1-p)}+X-1,
  \end{equation}
  signaling an instability toward ferromagnetism. The latter
  instability precedes the localization transition whenever
  $p>p_{\text{fm}}$, where
  \begin{equation}
    p_{\text{fm}}
    =
    \frac{4}{3}\left( 1+\frac{rX}{2(1-X)}\right) .
  \end{equation}
  These results reduce to the known values $p_{\text{fm}}$ $=$
  $\frac{4}{3}$ and $U_{\text{fm}}$ $=$ $(\sqrt{p/(p-1)}-1)$
  $U_{\text{c}}$ for $X=0$\mycite{.}{Brinkman70a,Vollhardt84a}
  
  Let us first consider the effect of the bond-charge interaction in
  the case of a symmetric DOS, $N(\epsilon)$ $=$ $N(-\epsilon)$, which
  results if hopping takes place only between different sublattices of
  a bipartite lattice. The Fermi energy at half-filling is then
  $\epsilon_{F}$ $=$ $0$, hence $r=0$. In this case $p_{\text{fm}}$
  $=$ $\frac{4}{3}$, the same criterion as for $X=0$. On the other
  hand, for an asymmetric DOS ferromagnetism is favored by $X>0$ if
  $r<0$, i.~e., $\epsilon_{F}<0$. This is the case if the Fermi energy
  is below the center of mass of the DOS, which means that there is
  large spectral weight below the Fermi energy. The tendency toward
  ferromagnetism in such a situation was already proposed long
  ago\mycite{.}{Hubbard63a,%
    Kanamori63a,Gutzwiller63a+64a+65a,Vollhardt99a}
  
\begin{figure}[h]
    \epsfxsize=\hsize
    \centerline{\epsfbox{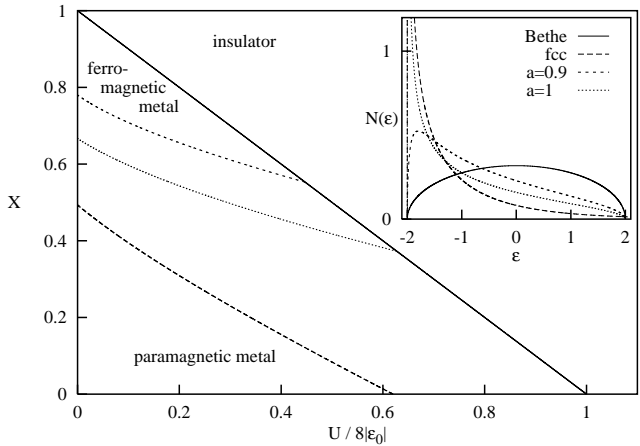}}
    \vspace*{3mm}
    \caption{Phase diagram for the Hubbard model with bond-charge
      interaction $X$ at half-filling ($n=1$). The Brink\-man-Rice
      metal-insulator transition takes place for $U_{\text{c}}(X)$ $=$
      $8|\protect\epsilon_0|(1-X)$ (solid line).  The dashed lines
      mark the ferromagnetic phase transition for the fcc lattice
      [Eq.~(\ref{fccdos})] and for the model DOS of Eq.~(\ref{tundos})
      for $a=1$ and $0.9$, respectively. The inset shows various
      densities of states, all with unit variance. The lower band-edge
      has been set to the same value for better comparison.
      \label{fig:phasediag}}
  \end{figure}

  We now consider nearest-neighbor hopping $t_{ij}$ $=$
  $-t^{\ast}/\sqrt{Z}$ and bond-charge interaction $X_{ij}$ $=$ $X$
  $t^{\ast}/\sqrt{Z}$ on several infinite-dimensional lattices.  Their
  densities of states and phase diagram are shown in
  Fig.~\ref{fig:phasediag}. The Gaussian DOS of the hypercubic
  lattice, $N(\epsilon)$ $=$ $\exp(-\epsilon^{2}/2)/\sqrt{2\pi}$, has
  a parameter $p$ $=$ $1/\pi$, and the semielliptic DOS of the
  Bethe lattice, $N(\epsilon)$ $=$ $\sqrt{4-\epsilon^2}/2\pi$, has
  $p=32/3\pi^{3}$, and both have $r=0$ due to particle-hole
  symmetry. Since $p<\frac{4}{3}$ in both cases, the metal-insulator
  transition at $U_{\text{c}}$, drawn as a solid line in the phase
  diagram in Fig.~\ref{fig:phasediag}, will mask the ferromagnetic
  phase. The variational phase diagram for $\hat{H}_{\text{Hubbard}}$
  (i.~e., $X$ $=$ $0$) on the hypercubic lattice was calculated by
  Fazekas et al.\mycite{,}{Fazekas90a} who predicted the
  ferromagnetic and
  antiferromagnetic phases to coexist as the system phase separates
  and to preempt the metal-insulator transition. Here we consider only
  homogeneous ferromagnetic phases, thus allowing the metal-insulator
  transition to take place, and do not attempt to distinguish between
  paramagnetic and ferromagnetic insulators at half-filling, which
  are degenerate in energy ($E=0$).

  A more complicated scenario arises if the lattice system is not
  particle-hole symmetric, so that the DOS is asymmetric and thus
  $r\neq 0$.  The generalized infinite-dimensional fcc lattice, with
  hopping scaled as $t_{ij}$ $=$ $-1/\sqrt{2D(D-1)}$, has a
  DOS (Ref.~\onlinecite{MuellerHartmann91a})
  \begin{equation}
    N(\epsilon)
    =
    \frac{\exp({-(1+\sqrt{2}\epsilon)/2})}{\sqrt{\pi (1+\sqrt{2}\epsilon)}}
    ,\label{fccdos}
  \end{equation}
  showing a square-root singularity at the lower band-edge. The
  Hubbard model (\ref{u-hamiltonian}) on this lattice has been studied
  numerically by Ulmke within the dynamical mean-field
  theory\mycite{,}{Ulmke98a} who found ferromagnetism at low enough
  temperatures and band-filling.  For half-filling one has
  $\epsilon_{F}$ $=$ $-0.3854$, $p$ $=$ $1.6157$, and $r=-1.0272$. Thus
  ferromagnetism occurs in this case already for $X=0$, but the
  critical $U$ is lowered by the presence of $X>0$. Hence a
  ferromagnetic phase is found for $U_{\text{fm}}(X)$ $<$ $U$ $< $
  $U_{\text{c}}(X)$.

  Finally, for the class of densities of states with $p$ $<$
  $\frac{4}{3}$ ferromagnetism is absent for $X=0$ for all $U$, and is
  only enabled by switching on the bond-charge interaction $X>0$. It
  is useful to consider a model DOS\mycite{,}{Wahle98a}
  \begin{equation}
    N(\epsilon)
    =
    \frac{1+\sqrt{1-a^{2}}}{2\pi}\frac{\sqrt{4-\epsilon ^{2}}}{2+a\epsilon}
    ,\label{tundos}
  \end{equation}
  where the tunable parameter $a$ $=$ $-1$ $\ldots$ $1$ determines the
  distribution of spectral weight: for $a=0$ this DOS reduces to the
  Bethe DOS, whereas for $a=1$ there is a square-root singularity at
  the lower band-edge similar to the fcc DOS. For $a=1$ we have $p$
  $=$ $1.1353$, $r$ $=$ $-0.5006$, while for $a=0.9$ the parameters
  are $p$ $=$ $1.1008$ and $r$ $=$ $-0.2821$. Thus in these cases only a
  metal-insulator transition is found for $X=0$, but for large enough
  $X$ a ferromagnetic phase is predicted, as shown in
  Fig.~\ref{fig:phasediag}.

\begin{figure}[h]
    \epsfxsize=\hsize
    \centerline{\epsfbox{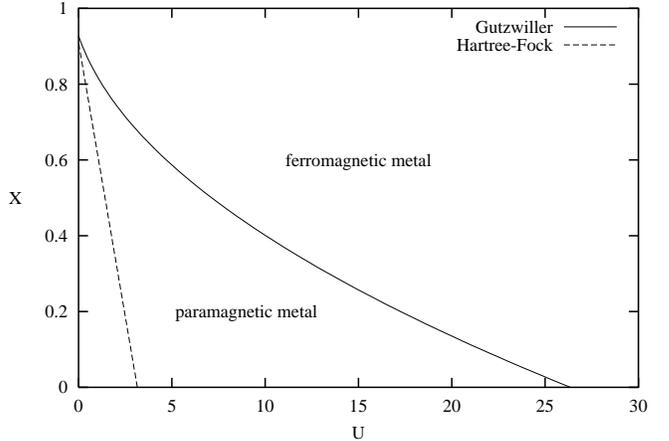}}
    \vspace*{3mm}
    \caption{Phase diagram for the Bethe lattice at density $n=0.9$.
      The Gutzwiller approximation gives a much smaller region of
      stability of ferromagnetism than Hartree-Fock
      theory.\label{fig:xbethe}}
  \end{figure}

  {\em Away from half-filling}.--- Since the metal-insulator
  transition takes place only at half-filling, metallic ferromagnetism
  occupies a larger part of the phase diagram for $n$ $\neq $ $1$.
  Figure~\ref{fig:xbethe} shows results for the Bethe lattice with
  particle density $n=0.9$. Whereas for half-filling the
  strong-coupling phase was insulating, now there is metallic behavior
  for all couplings, with ferromagnetism setting in for large $U$ when
  $X$ is small, and moving to small $U$ when $X$ becomes large.
  Compared to Hartree-Fock theory the Gutzwiller approximation
  predicts a much reduced region of stability of ferromagnetism, which
  is due to the correlated nature of the Gutzwiller wave function.
  While in Hartree-Fock theory, owing to the lack of correlations,
  double occupation can be reduced only through a global spin
  polarization of the system, the Gutzwiller wave function describes a
  paramagnetic state with reduced double occupation controlled by the
  variational parameter $g$.  Furthermore, the Hartree-Fock prediction
  of a phase boundary in a range where either $U$ or $X$ is comparable
  with the hopping amplitude (see Fig.~\ref{fig:xbethe}), is not
  consistent with the weak-coupling nature of this approximation,
  i.~e., self-consistent perturbation theory to {\em first} order in
  $U$ and $X$. Hence we expect that the Gutzwiller wave function in
  general provides a quantitatively better estimate than Hartree-Fock
  theory.
  
  \vspace*{\fill}

  \begin{figure}[h]
    \epsfxsize=\hsize
    \centerline{\epsfbox{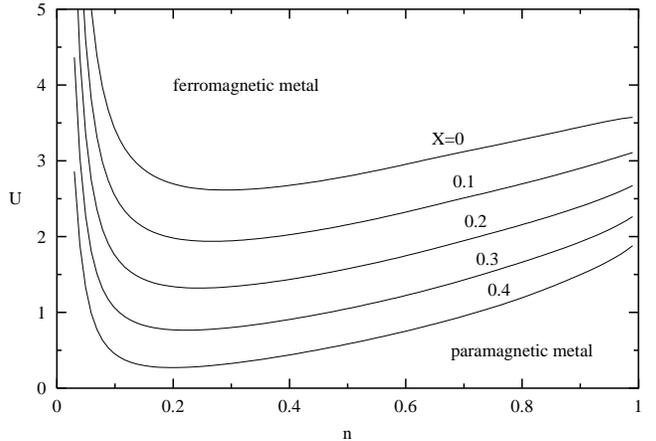}}
    \vspace*{3mm}
    \caption{Phase diagram for the model density of states
      [Eq.~(\ref{tundos})] for $a=0.9$ away from
      half-filling.\label{fig:xtunab}}
  \end{figure}
  \vspace*{8mm}

  Figure~\ref{fig:xtunab} shows the $U$ vs $n$ phase diagram for the
  model DOS of Eq.~(\ref{tundos}) with $a$ $=$ $0.9$ for several
  values of $X$.  Already for $X=0$ the Gutzwiller theory predicts a
  large region of ferromagnetic ground states, in qualitative
  agreement with numerical results\mycite{.}{Wahle98a} The bond-charge
  interaction again leads to a further stabilization of
  ferromagnetism.

  {\em Conclusion.}--- We found that within Gutzwiller's approach the
  bond-charge interaction can enhance the instability towards
  ferromagnetism both at and away from half-filling. This effect is
  particularly strong when the uncorrelated DOS is asymmetric and
  there is large spectral weight below the Fermi energy. This provides
  further support for the conclusion\mycite{}{Wahle98a,Vollhardt99a} that
  such a situation is favorable for ferromagnetism. At half-filling,
  the presence of the bond-charge interaction leads to a
  metal-insulator transition at lower values than in the standard
  Brinkman-Rice scenario, since it tends to immobilize the electrons.
  Although the Gutzwiller theory can be expected to be reliable only
  at small to intermediate couplings, it represents a major
  improvement over Hartree-Fock theory, which for example cannot
  describe a nonmagnetic localization transition.

  In conclusion the bond-charge
  interaction leads to a subtle competition between
  paramagnetism, ferromagnetism, and localization. Of course, a
  variational method is not capable of proving the actual stability of
  a phase. It can only provide estimates for the occurrence of
  instabilities.  Nevertheless, since the Gutzwiller theory treats
  kinetic and interaction effects nonperturbatively on the same
  footing, it provides additional insight into the physical mechanism
  behind these instabilities.

  This work was supported in part by the Sonderforschungsbereich 484
  of the Deutsche Forschungsgemeinschaft. We would like to thank K.
  Held for providing us with the Hartree-Fock data of
  Fig.~\ref{fig:xbethe}.


\begin{references}
\bibitem{footnote1}
    Although there is some ambiguity over what is called bond-charge
    interaction in the literature, we will adopt this name for $X_{ij}$
    throughout this paper.
\end{references}
\end{document}